# Vanadium Dioxide Thin Films Synthesized Using Low Thermal Budget Atmospheric Oxidation

Ashok P, Yogesh Singh Chauhan, and Amit Verma

Department of Electrical Engineering, Indian Institute of Technology Kanpur, Kanpur 208016, India

Email: ashok@iitk.ac.in and amitkver@iitk.ac.in

ABSTRACT: Vanadium dioxide is a complex oxide material, which shows large resistivity and optical reflectance change while transitioning from the insulator to metal phase at ~68 °C. In this work, we use a modified atmospheric thermal oxidation method to oxidize RF-sputtered Vanadium films. Structural, surface-morphology and phase-transition properties of the oxidized films as a function of oxidation duration are presented. Phase-pure $VO_2$ films are obtained by oxidizing ~130 nm Vanadium films in short oxidation duration of ~30 seconds. Compared to previous reports on $VO_2$ synthesis using atmospheric oxidation of Vanadium films of similar thickness, we obtain a reduction in oxidation duration by more than one order. Synthesized $VO_2$ thin film shows resistance switching of ~3 orders of magnitude. We demonstrate optical reflectance switching in long-wave infrared wavelengths in $VO_2$ films synthesized using atmospheric oxidation of Vanadium. The extracted refractive index of $VO_2$ in the insulating and in the metallic phase is in good agreement with $VO_2$ synthesized using other methods. The considerable reduction in oxidation time of $VO_2$ synthesis while retaining good resistance and optical switching properties will help in integration of $VO_2$ in limited thermal budget processes, enabling further applications of this phase-transition material.

## 1. Introduction

Vanadium dioxide ($VO_2$) is a transition metal oxide compound, which undergoes structural phase transition at ~68°C from insulating monoclinic phase to metallic tetragonal phase [1,2]. In single crystal $VO_2$, insulator to metal transition (IMT) is accompanied by nearly five orders change in resistivity and a substantial change in infrared optical reflectance [3]. These properties make $VO_2$ a suitable candidate for applications such as ultra-thin absorbers [4], thermo chromic films [5-8], memristors [9], selectors for resistive random-access memory [10], capacitive memory [11], chemical sensors [12], oscillators [13], and micro actuators [14]. Recently, IMT property of $VO_2$ has also been exploited to break the Boltzmann limit of current switching in field effect transistors [15,16].

All the potential applications of $VO_2$ depend on thin film form of this phase change material. In the past, various methods have been used to deposit $VO_2$ thin films. These include reactive sputtering [17-19], chemical vapor deposition (CVD) [20-22], sol-gel synthesis [23], pulsed laser deposition (PLD) [24, 25], reactive-evaporation [26], molecular beam epitaxy (MBE) [27,28], atomic layer deposition [29], ion implantation [30], and thermal oxidation of Vanadium [31-41]. Large resistive switching ratio in $VO_2$ deposited on sapphire substrate has been reported using these methods: PLD (~5 orders) [24], CVD (~4 orders) [22], MBE (~4 orders) [28], sol-gel (~4orders) [23], sputtering (~4.5 orders) [19]. Growth of single-phase $VO_2$ is quite challenging due to multiple valence states of Vanadium (V). The V-O system exhibits at least 22 stable phases such as $V_2O_5$, $V_2O_3$, $V_3O_5$, $V_6O_{13}$ etc [42, 43]. Almost all thin film deposition methods for $VO_2$ require stringent control of oxygen partial pressure during the deposition to obtain phase-pure material. Even a slight variation in oxygen partial pressure leads to different or mixed phases in the grown thin film [17, 18].

$VO_2$ synthesis by thermal oxidation of V metallic films can be done by either vacuum thermal oxidation in controlled oxygen environment [39] or atmospheric pressure thermal oxidation (APTO) [36]. APTO method is inherently simple as it does not require vacuum or precise gas flow/pressure control during oxidation. Atmosphere has ~21% oxygen concentration, which is directly used for oxidizing V films. The oxidation time and oxidation temperature are the most crucial parameters for obtaining phase-pure $VO_2$ thin films using APTO method. Typical oxidation durations are 5-30 minutes [36-39] (depending on V film thickness) and oxidation temperatures are 420-470 °C [36-39]. In most reported articles, DC sputtered V thin films have been used for oxidation [36-39]. Though, more than three orders of electrical switching has been reported with this synthesis method [38], long wave infrared (LWIR) optical reflectance switching of $VO_2$ has yet not been demonstrated in APTO $VO_2$ films. Surface morphology evolution of V as a function of oxidation duration has also not been studied.

In this work, we have synthesized phase-pure $VO_2$ thin films on c-plane Sapphire substrates using APTO of radio-frequency (RF) sputtered V thin films. We performed the oxidation on a hot plate followed by quenching to have a step temperature profile during the oxidation process. Unlike previous studies on APTO of Vanadium, the step temperature profile used in this study allows precise control of the oxidation time and hence oxygen content in the films. The structural composition as a function of oxidation time was characterized by X-ray diffraction (XRD) and Raman spectroscopy. Morphological characterization of all samples was done by Field Emission Scanning Electron Microscope (FE-SEM) and Atomic Force Microscope (AFM). Temperature dependent four-probe electrical resistance and Fourier-Transform Infrared spectroscopy (FTIR) optical reflectance measurements were used to study the IMT properties of $VO_2$ as a function of oxidation time. Complex refractive index of $VO_2$ in insulating and metallic states was extracted from reflectance measurements. By varying the oxidation duration, we were able to obtain phase-pure $VO_2$ thin films for ultra-short oxidation duration of ~30 seconds. This oxidation duration is more than one order less compared to previous APTO reports of similar thickness of V films [36-39]. This film shows ~3 orders of resistance switching and up to 60% optical reflectance switching in LWIR frequencies.

## 2. Experimental Procedure

To synthesize Vanadium oxide thin films, metallic V was deposited on c-plane Sapphire substrates using RF magnetron sputtering. Prior to deposition, c-plane Sapphire substrate was cleaned using acetone and iso-propyl alcohol in a sonicator, and blow dried using dry nitrogen. A 2-inch diameter V sputtering target of 99.9% purity was used.

The sputtering chamber was pumped to a base pressure of 5 x10$^{-4}$ Pa. During the deposition, Argon gas flow rate of 30 sccm was introduced to maintain the chamber pressure at 1 Pa, RF power was maintained at 90 W, and the substrate was kept at room temperature. For 90 minutes deposition, V film thickness of ~130 nm was obtained as measured using a KLA-Tencor stylus profilometer.

To perform atmospheric oxidation experiment, the sample was diced into pieces and put on a hot plate maintained at 450 °C (as measured using the hot plate built-in thermocouple) and kept in ambient air atmosphere. To precisely end the oxidation, we quenched the sample by putting it on a cold plate. The heating and cooling rates of the sample were measured using an optical thermometer and found to be greater than 100 °C/s and 110 °C/s, respectively. Samples with oxidation duration ($T_{oxd}$) of 15-360 seconds were prepared. The sample preparation and oxidation process is shown in Fig. 1(a). Optical images of all the samples are shown in Fig 1(b). Structural characterization of the oxidized samples was done using X-ray diffraction (Rigaku miniflex pro) using Cu K$\alpha$ radiation and Raman spectroscopy was done using an excitation laser of wavelength 532 nm (Acton research corporation spectra pro 2500i). Surface morphological characterization was studied using FE-SEM (Jeol model JSM-7100F with accelerating voltage of 10 kV) and AFM (Oxford Instruments MFP-3D). Temperature dependent resistance measurements were carried out using a four-point probe set up and a temperature controlled oven. Infrared optical reflectance was measured by an IR microscope (Agilent, Cary 600) connected to the FTIR spectrometer (Agilent, Cary 660).

## 3. Results and Discussion

### 3.1 Structural Characterization

Fig.2(a) shows $\theta$-2$\theta$ X-ray diffraction scans of the thermally oxidized V thin films of different $T_{oxd}$ durations. There is no diffraction peak for unoxidized V thin film ($T_{oxd}$ = 0 s). This absence of diffraction peak is due to the amorphous nature of room temperature deposited V. The films with $T_{oxd}$ = 15-180s show a diffraction peak around 39.74°, which corresponds to monoclinic (020) planes of VO$_2$ [24, 25]. This peak indicates the oriented growth of VO$_2$ thin films on the c-plane Sapphire substrate [24]. The intensity of the (020) diffraction peak is highest for samples with $T_{oxd}$ = 30 s and $T_{oxd}$ = 45 s, suggesting higher VO$_2$ content in these samples. For $T_{oxd}$> 45 s, (020) diffraction peak intensity starts reducing, suggesting decrease in VO$_2$ content of the film. There is no diffraction peak of VO$_2$ in $\theta$-2$\theta$ scan of $T_{oxd}$ = 240-360 s samples.

Raman spectrum of oxidized V thin films is shown in Fig.2(b). Films with $T_{oxd}$ =15 s, 30 s & 45 s show strong VO$_2$ peaks in the Raman spectrum. While $T_{oxd}$=15 s and $T_{oxd}$=30 s samples show only VO$_2$ peaks, $T_{oxd}$=45 s film shows a weak V$_2$O$_5$ peak around 148 cm$^{-1}$. Higher oxidation durations show increasingly dominant V$_2$O$_5$ peaks. The intensity of 148 cm$^{-1}$ V$_2$O$_5$ Raman peak was found to monotonically increase with increasing oxidation duration, suggesting increase in V$_2$O$_5$ content of the films. $T_{oxd}$ = 45–240 s films show a mixed phase of VO$_2$ and V$_2$O$_5$ in the Raman spectrum, whereas $T_{oxd}$= 300–360 s films show only V$_2$O$_5$ peaks [44-46]. From the structural characterization, it was found that during the initial stages of oxidation, only VO$_2$ forms followed by V$_2$O$_5$. This observation agrees well with previous vanadium oxidation studies [33, 39, 52], which attribute this behavior to the kinetics of the oxidation process [33, 39, 52].

## 3.2 Surface Morphology Characterization

Fig. 3(a) and Fig. 3(b) show FE-SEM and AFM images of all the oxidized samples, respectively. Fig. 3(c) shows the measured root mean square (RMS) roughness of the oxidized samples as a function of $T_{oxd}$. A gradual surface roughening is observed for initial oxidation duration of $T_{oxd} = 15\text{-}45$ s as crystallite grain size increases with increasing $T_{oxd}$. After 45 s oxidation duration, there is a steep drop in surface roughness for $T_{oxd} = 60$ s. This decrease in roughness is possibly due to increasingly dominant $V_2O_5$ content of the sample (as observed from the structural characterization). $V_2O_5$ decomposes around 400ºC in vacuum [50]. In our atmospheric oxidation experiments, some $V_2O_5$ thermal decomposition can be expected as $V_2O_5$ content in the film increases which can lead to the dip in surface roughness. Beyond $T_{oxd} = 120$ s, the grain size is observed to saturate possibly due to a balance between the $V_2O_5$ grain growth and dissociation. Gradual roughening of the films is however observed with increasing oxidation duration ($T_{oxd} > 120$ s) as pits are formed on the films which can be clearly observed in the corresponding AFM images. This pit formation in the films is possibly due to $V_2O_5$ decomposition also.

## 3.3 Electrical Characterization

As mentioned earlier, $VO_2$ resistivity changes by multiple orders as it undergoes an IMT across ~68°C. Fig. 4 (a) summarizes the resistance switching ratio $R_{30\,ºC}/R_{110ºC}$ of the samples as a function of $T_{oxd}$. $T_{oxd} = 15$ s film shows more than one order of resistance switching. From the structural characterization, this sample shows only $VO_2$ phase, but the relatively small resistance switching suggests only partial oxidation of V in this sample. Among all samples, $T_{oxd} = 30$ s film shows best resistance switching ratio of nearly three orders of magnitude (Fig. 4(b) shows the temperature-dependent resistance measurement of this sample). Coupled with structural characterization data, we can conclude that this sample is almost phase pure $VO_2$ with no other Vanadium oxide or unoxidized V content. With an increase in $T_{oxd}$, $V_2O_5$ content of the films increases, leading to reduction in IMT resistance switching and ultimately leading to no sharp switching ($T_{oxd} = 240$ s, 300 s, 360 s) in the temperature range of measurement.

Fig. 4 (c) shows IMT properties such as transition temperature $T_{IMT}$ (during heating scan), hysteresis in IMT & MIT scans $\Delta T_{Hys}$, and transition interval $\Delta T_{Tr}$ (during heating scan) of all the samples. To extract above mentioned quantities, the derivative of $\log_{10}R(T)$ is plotted and fitted with a Gaussian to precisely determine $T_{IMT}$, $\Delta T_{Hys}$ and $\Delta T_{Tr}$ [49]. $T_{oxd} = 30$ s film shows transition temperature around 69 ºC and hysteresis width of 7 ºC. Activation energy of $VO_2$ in the insulating phase was extracted to be 0.56 eV from Fig. 4(b). This value is quite close to activation energy in bulk $VO_2$ [51]. To test the endurance of the $T_{oxd} = 30$ s sample with thermal cycling across the phase-transition temperature, we performed 10 resistance vs temperature measurements as shown in Fig. 4(d). No significant deterioration in the resistance switching properties of the films was observed suggesting that the synthesized films are suitable for device applications.

### 3.4 Optical Characterization

As $VO_2$ undergoes IMT, its optical reflectance in infrared wavelengths is expected to change drastically [4]. Fig. 5(a) shows optical reflectance switching at 9.6 µm for all the oxidized samples. $T_{oxd}$=15 s and 30 s films show reflectance switching of almost 60%. Fig. 5(b) shows the room temperature and high temperature (100 °C) reflectance measurement of $T_{oxd}$=30 s sample and Lorentz-Drude oscillator fit of the same. Large reflectance switching is obtained in the ~ 7-11 µm LWIR wavelength window with a peak reflectance switching of ~ 60% at 9.6 µm. For $T_{oxd}$=45 s and 60 s films, the switching reduces to 52% and 50%, respectively. Samples with $T_{oxd} \geq 240$ s do not exhibit any significant optical switching as expected from samples with dominant $V_2O_5$ content.

To extract the complex refractive index (n,k) of the $VO_2$ film in insulating and metallic states, we assumed a Lorentz-Drude oscillator form for the dielectric constant [48].

$$\varepsilon(\omega) = \varepsilon_\infty - \frac{\omega_p^2}{\omega(\omega + i\omega_c)} + \sum_j \frac{f_j}{1 - \frac{\omega^2}{\omega_j^2} - i\gamma_j \omega/\omega_j} \qquad (1)$$

Where $\omega_p$ is plasma frequency, $\omega_c$ is collision frequency, $\omega_j$ is phonon resonance frequency, $f_j$ is strength of oscillator, and $\gamma_j$ is the line width. To find the oscillator parameters, we calculated the reflectance spectra and performed a fit to the measured reflectance spectra. For insulating $VO_2$ phase, we used two Lorentz oscillators whereas an additional Drude term was added for the metallic phase. The optimized oscillator parameters in the insulating and in the metallic phase are as shown in Table 1. From the Drude term in metallic $VO_2$ phase, the extracted plasma frequency agrees well with previous reports [48]. Fig. 5 (c,d) show a comparison of extracted refractive index for $T_{oxd}$=30 s APTO $VO_2$ sample with published refractive indices of $VO_2$ deposited by other methods such as sputtering and sol-gel [47]. In the LWIR wavelengths, the real part (n) of the $VO_2$ refractive index is found to increase by 7-9 in the metallic phase compared to the insulating phase. In the insulating phase, the extracted extinction coefficient (k) is almost zero (2-10 µm), which shows the transparent behavior of $VO_2$ in IR wavelengths. Extracted n and k values of $VO_2$ insulating phase agrees well with the reported values of $VO_2$ films prepared using both sputtering and sol-gel methods [47]. However, in the metallic $VO_2$ phase, only n values agree with $VO_2$ films prepared using both deposition methods. The k values match with $VO_2$ films prepared using sol-gel method only. This behavior is possibly due to smaller crystallite size of both sol-gel and APTO $VO_2$ films compared to the films deposited using sputtering at high-substrate temperatures.

### 4. Conclusion

In this work, we have demonstrated phase pure $VO_2$ thin films by simple atmospheric oxidation of V thin films deposited using RF magnetron sputtering. For the oxidation, we have used a step temperature profile for precise control over oxygen content of the films, thus enabling good resistance (~3 orders) and optical reflectance switching (~60% at 9.6 µm) across the $VO_2$ phase transition. Large optical reflectance switching in the long wave infrared wavelengths is shown in $VO_2$ films synthesized using APTO. The extracted complex refractive index for APTO synthesized $VO_2$ agrees well with refractive index of $VO_2$ synthesized using other deposition methods. Compared to previous APTO studies, we have been able to reduce the oxidation time by more than one order. This significant reduction in the oxidation time can enable $VO_2$ integration in processes with limited thermal budget, opening further avenues of applications of this material.

**Acknowledgements**

This project was supported by IIT Kanpur initiation grant and used Advanced Center for Materials Science (ACMS) XRD facility and Materials Science and Engineering (MSE) Raman characterization facility.

## REFERENCES

[1] F. J. Morin, Oxides which show a metal-to-insulator transition at the Neel temperature, Phys. Rev. Lett. 3 (1959) 34–36.

[2] J. B. Goodenough, The Two Components of the Crystallographic Transition in $VO_2$, J. Solid State Chem. 3 (1971) 490–500.

[3] C. Lamsal, N.M. Ravindra, Optical properties of vanadium oxides-an analysis, J. Mater. Sci. 48 (2013) 6341–6351.

[4] M. A. Kats, D. Sharma, J. Lin, P. Genevet, R. Blanchard, Z. Yang, M.M. Qazilbash, D.N. Basov, S. Ramanathan, F. Capasso, Ultra-thin perfect absorber employing a tunable phase change material, Appl. Phys. Lett. 101 (22) (2012) 221101.

[5] Y. Gao, H. Luo, Z. Zhang, L. Kang, Z. Chen, J. Du, M. Kanehira, C. Cao, Nanoceramic $VO_2$ thermochromic smart glass: A review on progress in solution processing, NanoEnergy,1 (2012) 221–246.

[6] D. Louloudakis, D. Vernardou, E. Spanakis, N. Katsarakis, E. Koudoumas, Thermochromic vanadium oxide coatings grown by APCVD at low temperatures, Phys. Procedia. 46 (2013) 137–141.

[7] D. Vernardou, M.E. Pemble, D.W. Sheel, In-situ Fourier transform infrared spectroscopy gas phase studies of vanadium (IV) oxide coating by atmospheric pressure chemical vapour deposition using vanadyl (IV) acetylacetonate, Thin Solid Films. 516 (2008) 4502–4507.

[8] D. Louloudakis, D. Vernardou, E. Spanakis, S. Dokianakis, M. Panagopoulou, G. Raptis, E. Aperathitis, G. Kiriakidis, N. Katsarakis, E. Koudoumas, Effect of $O_2$ flow rate on the thermochromic performance of $VO_2$ coatings grown by atmospheric pressure CVD, Phys. Status Solidi C. 12 (2015) 856–860.

[9] T. Driscoll, H.T. Kim, B.G. Chae, M. Di Ventra, D.N. Basov, Phase-transition driven memristive system, Appl. Phys. Lett. 95 (4) (2009) 043503.

[10] I. Radu, B. Govoreanu, K. Martens, Vanadium Dioxide for Selector Applications, ECS Trans. 58 (2013) 249–258.

[11] S. Ramanathan, S. Tiwari, A new single element phase transition memory, Proc. 10th IEEE Conf. Nanotechnol. (2010) 439–442.

[12] E. Strelcov, Y. Lilach, A. Kolmakov, Gas Sensor Based on Metal-Insulator Transition in $VO_2$ Nanowire Thermistor, Nano Lett. 9 (2009) 2322–2326.

[13] Y.W. Lee, B.J. Kim, J.W. Lim, S.J. Yun, S. Choi, B.G. Chae, G. Kim, H.T. Kim, Metal-insulator transition-induced electrical oscillation in vanadium dioxide thin film, Appl. Phys. Lett. 92 (16) (2008) 161903.

[14] K. Liu, S. Lee, S. Yang, O. Delaire, J. Wu, Recent progresses on physics and applications of vanadium dioxide, Mater. Today, 21 (2018) 875-896.

[15] N. Shukla, A. V Thathachary, A. Agrawal, H. Paik, A. Aziz, D. G. Schlom, S. K. Gupta, R. Engel-Herbert, S. Datta, A steep-slope transistor based on abrupt electronic phase transition, Nat. Commun. 6 (2015) 8475.

[16] A. Verma, B. Song, B. Downey, V.D. Wheeler, D. J. Meyer, H. G. Xing, D. Jena, Steep Sub-Boltzmann Switching in AlGaN/GaN Phase-FETs with ALD $VO_2$, IEEE Trans. Electron Devices. 65 (2018) 945–949.

[17] E. Kusano, J. A. Theil, J.A. Thornton, Deposition of vanadium oxide films by direct-current magnetron reactive sputtering, J. Vac. Sci. Technol. A. 6 (1988) 1663–1667.

[18] D. Ruzmetov, S. D. Senanayake, V. Narayanamurti, S. Ramanathan, Correlation between metal-insulator transition characteristics and electronic structure changes in vanadium oxide thin films, Phys. Rev. B. 77 (19) (2008) 195442.

[19] Y. Zhao, J. Hwan Lee, Y. Zhu, M. Nazari, C. Chen, H. Wang, A. Bernussi, M. Holtz, Z. Fan, Structural, electrical, and terahertz transmission properties of $VO_2$ thin films grown on c-, r-, and m-plane sapphire substrates, J. Appl. Phys. 111 (2012) 053533.

[20] H. Zhane, H.L.M. Chans, J. Guo, T.J. Zhane, Microstructure of epitaxial $VO_2$ thin films deposited on (1120) sapphire by MOCVD, J. Mater. Res. 9 (1994) 2264–2271.

[21] M.B. Sahana, G.N. Subbanna, S.A. Shivashankar, Phase transformation and semiconductor-metal transition in thin films of $VO_2$ deposited by low-pressure metalorganic chemical vapor deposition, J. Appl. Phys. 92 (11) (2002) 6495–6504.

[22] T. Maruyama, Y. Ikuta, Vanadium dioxide thin films prepared by chemical vapour deposition from vanadium (III) acetylacetonate, J. Mater. Sci., 28 (1993), pp. 5073-5078

[23] B. Chae, H. Kim, S. Yun, B. Kim, Y. Lee, D. Youn, K. Kang, Highly oriented $VO_2$ thin films prepared by Sol-Gel deposition, Electrochem. Solid-State Lett. 9 (2006) C12–C14.

[24] D.H. Kim, H.S. Kwok, Pulsed laser deposition of $VO_2$ thin films, Appl. Phys. Lett. 65 (25) (1994) 3188–3190.

[25] T. H. Yang, R. Aggarwal, A. Gupta, H. Zhou, R.J. Narayan, J. Narayan, Semiconductor-metal transition characteristics of $VO_2$ thin films grown on c-and r-sapphire substrates, J. Appl. Phys. 107 (5) (2010) 053514.

[26] F. C. Case, Low temperature deposition of $VO_2$ thin films, J. Vac. Sci. Technol. A. 8 (1990) 1395.

[27] H. Paik, J. A. Moyer, T. Spila, J.W. Tashman, J. A. Mundy, E. Freeman, N. Shukla, J. M. Lapano, R. Engel-Herbert, W. Zander, J. Schubert, D. A. Muller, S. Datta, P. Schiffer, D.G. Schlom, Transport properties of ultra-thin $VO_2$ films on (001) $TiO_2$ grown by reactive molecular-beam epitaxy, Appl. Phys. Lett. 107 (16) (2015) 163101.

[28] L. L. Fan, S. Chen, Y. F. Wu, F. H. Chen, W. S. Chu, X. Chen, C. W. Zou, Z. Y. Wu, Growth and phase transition characteristics of pure M-phase VO2 epitaxial film prepared by oxide molecular beam epitaxy, Appl. Phys. Lett. 103 (2013).

[29] M. J. Tadjer, V. D. Wheeler, B. P. Downey, Z. R. Robinson, D. J. Meyer, C. R. Eddy, F. J. Kub, Temperature and electric field induced metal-insulator transition in atomic layer deposited $VO_2$ thin films, Solid. State. Electron. 136 (2017) 30–35.

[30] R. Lopez, L. A. Boatner, T. E. Haynes, L. C. Feldman, R. F. Haglund, Synthesis and characterization of size-controlled vanadium dioxide nanocrystals in a fused silica matrix, J. Appl. Phys. 92 (7) (2002) 4031–4036.

[31] I. Balberg, S. Trokman, High-contrast optical storage in $VO_2$ films, J. Appl. Phys. 46 (5) (1975) 2111–2119.

[32] F. C. Case, Simple resistance model fit to the oxidation of a vanadium film into $VO_2$, J. Vac. Sci. Technol. A Vacuum, Surfaces, Film. 6 (1988) 123–127.

[33] A. Mukherjee, S. P. Wach, An investigation of the kinetics and stability of $VO_2$, J. Less-Common Met. 132 (1987) 107–113.

[34] S. Jiang, C. Ye, M. S. R. Khan, C. Granqvist, Evolution of thermochromism during oxidation of evaporated vanadium films, Appl. Opt. 30 (1991) 847–851.

[35] M. Gurvitch, S. Luryi, A. Polyakov, A. Shabalov, M. Dudley, G. Wang, S. Ge, V. Yakovlev, $VO_2$ films with strong semiconductor to metal phase transition prepared by the precursor oxidation process, J. Appl. Phys. 102 (3) (2007) 033504.

[36] X. Xu, A. Yin, X. Du, J. Wang, J. Liu, X. He, X. Liu, Y. Huan, A novel sputtering oxidation coupling (SOC) method to fabricate $VO_2$ thin film, Appl. Surf. Sci. 256 (2010) 2750–2753.

[37] X. Xu, X. He, G. Wang, X. Yuan, X. Liu, H. Huang, S. Yao, H. Xing, X. Chen, J. Chu, The study of optimal oxidation time and different temperatures for high quality $VO_2$ thin film based on the sputtering oxidation coupling method, Appl. Surf. Sci. 257 (2011) 8824–8827.

[38] X. Xu, X. He, H. Wang, Q. Gu, S. Shi, H. Xing, C. Wang, J. Zhang, X. Chen, J. Chu, The extremely narrow hysteresis width of phase transition in nanocrystalline $VO_2$ thin films with the flake grain structures, Appl. Surf. Sci. 261 (2012) 83–87.

[39] G. Rampelberg, B. De Schutter, W. Devulder, K. Martens, I. Radu, C. Detavernier, In situ X-ray diffraction study of the controlled oxidation and reduction in the V–O system for the synthesis of $VO_2$ and $V_2O_3$ thin films, J. Mater. Chem. C. 3 (2015) 11357–11365.

[40] C. Ba, S. T. Bah, M. D'Auteuil, P. V Ashrit, R. Vallée, Fabrication of high-quality $VO_2$ thin films by ion-assisted dual ac magnetron sputtering, ACS Appl. Mater. Interfaces. 5 (2013) 12520–12525.

[41] S. Yu, S. Wang, M. Lu, L. Zuo, A metal-insulator transition study of $VO_2$ thin films grown on sapphire substrates, J. Appl. Phys. 122 (23) (2017) 235102.

[42] J. Nag, R. F. Haglund, Synthesis of vanadium dioxide thin films and nanoparticles, J. Phys. Condens. Matter. 20 (26) (2008) 264016.

[43] Y. Kang, Critical evaluation and thermodynamic optimization of the $VO–VO_{2.5}$ system, J. Eur. Ceram. Soc. 32 (2012) 3187–3198.

[44] R. Srivastava, L. L. Chase, Raman spectrum of semiconducting and metallic $VO_2$, Phys. Rev. Lett. 27 (1971) 727–730.

[45] J. C. Parker, Raman scattering from $VO_2$ single crystals: A study of the effects of surface oxidation, Phys. Rev. B. 42 (1990) 3164–3166.

[46] A. G. S. Filho, O. P. Ferreira, E. J. G. Santos, J. M. Filho, O. L. Alves, Raman spectra in vanadate nanotubes revisited, Nano Lett. 4 (2004) 2099–2104.

[47] C. Wan, Z. Zhang, D. Woolf, C.M. Hessel, J. Rensberg, J.M. Hensley, Y. Xiao, A. Shahsafi, J. Salman, S. Richter, Y. Sun, M.M. Qazilbash, R. Schmidt-Grund, C. Ronning, S. Ramanathan, M. A. Kats, Optical Properties of Thin-Film Vanadium Dioxide from the Visible to the Far Infrared, Ann. Phys. 531 (2019) 1900188.

[48] B. Rajeswaran, J. K. Pradhan, S. Anantha Ramakrishna, A. M. Umarji, Thermochromic $VO_2$ thin films on ITO-coated glass substrates for broadband high absorption at infra-red frequencies, J. Appl. Phys. 122 (16) (2017) 163107.

[49] D. Ruzmetov, K. T. Zawilski, V. Narayanamurti, S. Ramanathan, Structure-functional property relationships in rf-sputtered vanadium dioxide thin films, J. Appl. Phys. 102 (11) (2007) 113715.

[50] D. S. Su, R. Schlögl, Thermal decomposition of divanadium pentoxide $V_2O_5$: Towards a nanocrystalline $V_2O_3$ phase, Catal. Letters. 83 (2002) 115–119.

[51] A. Zylbersztejn, N. F. Mott, Metal-insulator transition in vanadium dioxide, Phys. Rev. B. 11 (1975) 4383–4394.

[52] A. Mukherjee, S. P. Wach, Kinetics of the oxidation of vanadium in the temperature range 350-950 ℃, J. Less-Common Met. 92 (1983) 289–300.

Table 1. Lorentz- Drude oscillator parameters of APTO VO$_2$ film (thickness = 280 nm) in both insulating and metallic phase.

|  | $\varepsilon_\infty$ | $\hbar\omega_p$ (eV) | $\hbar\omega_c$ (eV) | $\hbar\omega_1$ (eV) | $\hbar\omega_2$ (eV) | $\gamma_1$ | $\gamma_2$ | $f_1$ | $f_2$ |
|---|---|---|---|---|---|---|---|---|---|
| Insulating phase | 7.756 | - | - | 0.0011 | 0.065 | 12.92 | 0.1988 | 3.274 | 9.594 |
| Metallic phase | 4.42 | 2.61 | 1.74 | 0.1639 | 0.1654 | 1.835 | 0.6785 | 86.75 | 2.44 |

# FIGURES

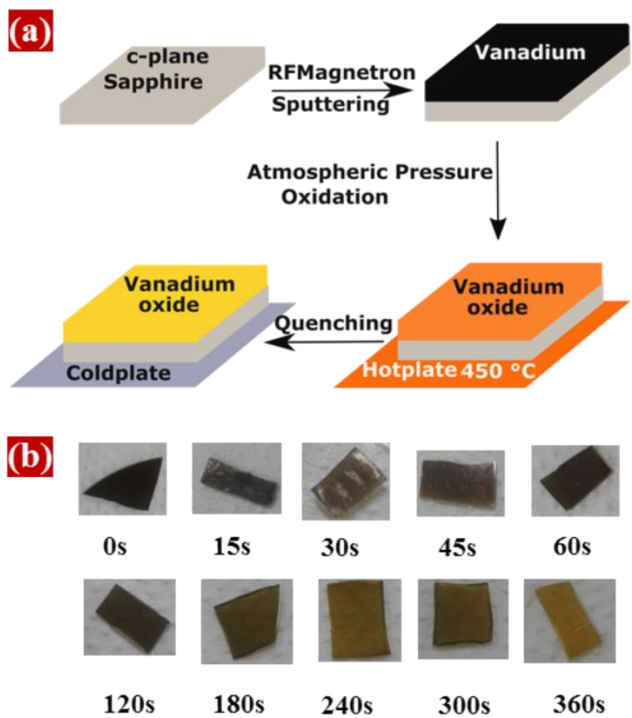

FIG. 1. (a) Atmospheric pressure thermal oxidation (APTO) process used to synthesize Vanadium oxide thin films. (b) Optical images of all oxidized samples. The oxidation duration is mentioned below each image.

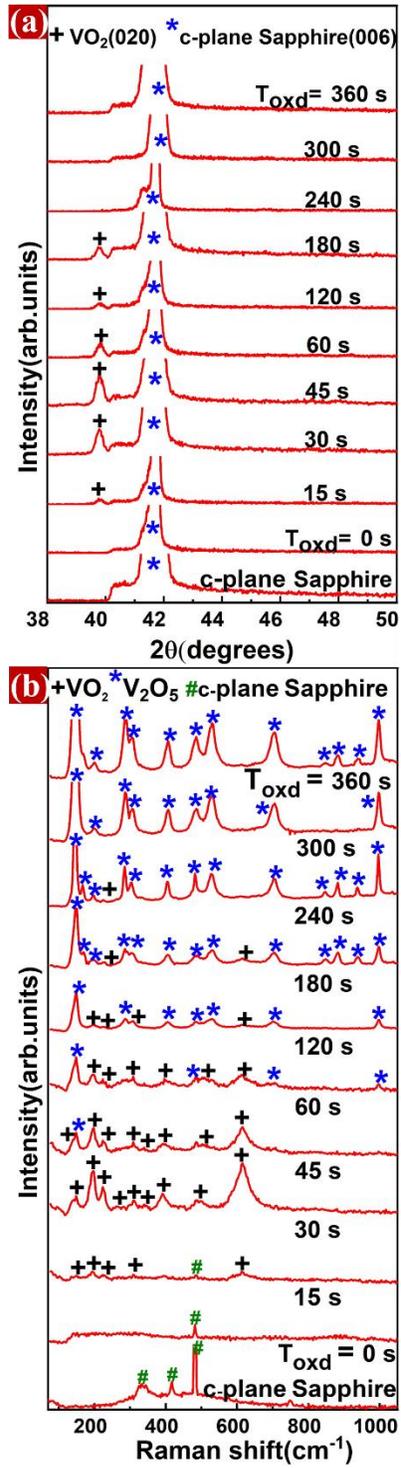

FIG. 2.(a) X-ray diffraction, (b)Raman spectra of oxidized Vanadium thin films as a function of oxidation time $T_{oxd}$.

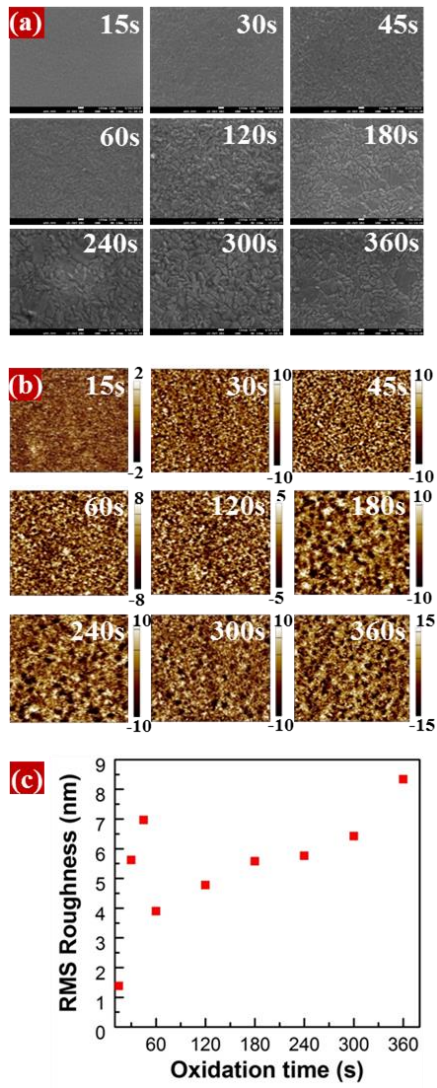

FIG. 3.(a) FESEM images of all the oxidized samples (Oxidation time $T_{oxd}$ is mentioned in the inset of each image). All FESEM images have the same magnification and scale bar of 100 nm,(b) AFM images of all the oxidized samples (Oxidation time $T_{oxd}$ is mentioned in the inset of each image and scale bar in nm is shown on the side of each image). All images have the same scanned area of 10 μm x10 μm.(c) Surface RMS roughness measured using AFM as a function of oxidation time.

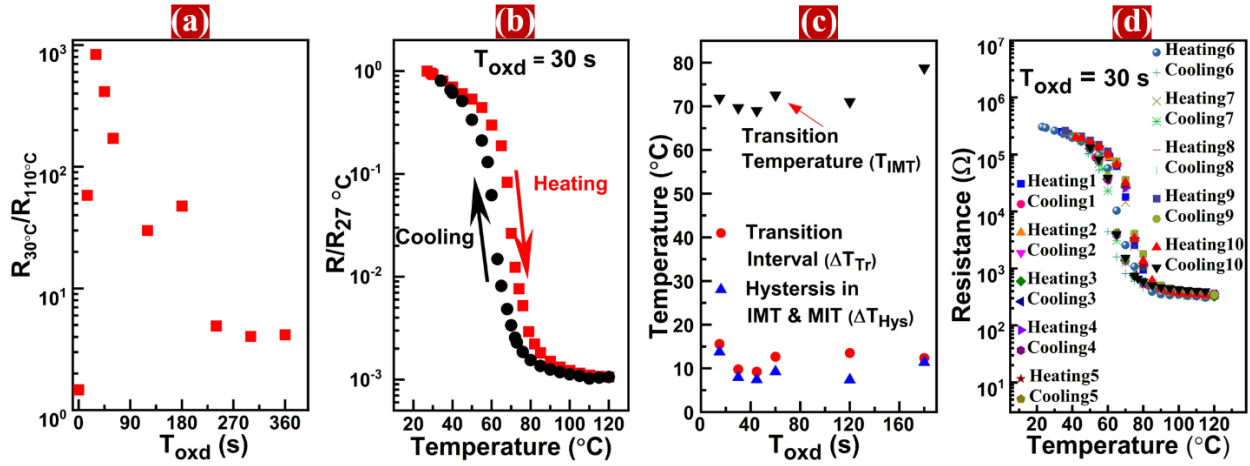

FIG. 4. (a) Variation of four-probe resistance switching as a function of oxidation duration showing highest switching for $T_{oxd} = 30$ s (b) Resistance switching in $T_{oxd} = 30$ s sample showing ~ 3 orders resistance switching with temperature. (c) IMT properties such as transition temperature (while heating), transition width (while heating) and hysteresis width as a function of oxidation time $T_{oxd}$ (sample with $T_{oxd} > 180$ s do not show IMT resistance switching). (d) Thermal cycling data (10 cycles) for $T_{oxd} = 30$ s sample showing no deterioration in resistance switching after repetitive thermal cycling.

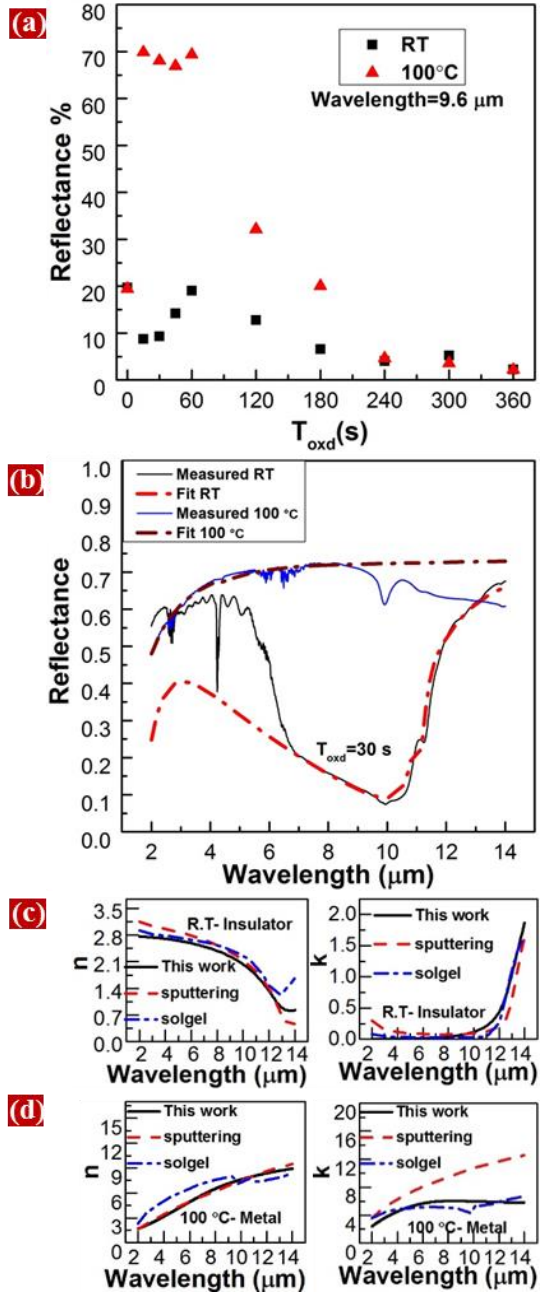

FIG. 5. (a) Optical reflectance switching (between RT and 100 °C) of oxidized samples at 9.6 µm as a function of oxidation time $T_{oxd}$ showing best switching for $T_{oxd}$ = 15 s & 30 s (b) FTIR reflectance spectrum of $T_{oxd}$= 30 s sample measured at room temperature and 100 °C showing large optical reflectance switching in long wave infra-red wavelengths; a Lorentz-Drude fit of the reflectance spectra is also shown. (c,d) Comparison of $VO_2$ complex refractive index (n,k) in the insulating and metallic phase as found in this work with the published results for $VO_2$ deposited by other methods.